\documentstyle[12pt,epsf]{article}

\def\alt{\;\raise.2ex\hbox{$<$}\hskip-.8em\lower.9ex\hbox{$\sim$}\;}

\begin{document}

\font\fortssbx=cmssbx10 scaled \magstep2
\hbox to \hsize{
\hbox{\fortssbx University of Wisconsin - Madison}
\hfill$\vcenter{\hbox{\bf MADPH-97-983}
                \hbox{January 1997}}$ }

\vspace{.5in}

\begin{center}
{\large\bf The anomalous threshold, confinement, and an essential singularity in the heavy-light form factor}\\[5mm]
M. G. Olsson\\
\it Physics Department, University of Wisconsin, Madison, WI 53706
\end{center}

\vspace{1in}

\begin{abstract}
The analytic behavior of the heavy-light meson form factor is investigated using several relativistic examples including unconfined, weakly confined, and strongly confined mesons. It is observed that confinement erases the anomalous threshold singularity and also induces an essential singularity at the normal annihilation threshold. In the weak confinement limit, the {\em would be} anomalous threshold contribution is identical to that of the real singularity on its space-like side.
\end{abstract}

\thispagestyle{empty}
\newpage

\section{Introduction}

As discussed in the seminal work of Jaffe and and Mende\cite{jaffe} there are circumstances where the constituent nature of a meson is of great importance and an effective theory (especially using dispersion relations) is awkward at best. These authors point out that although confinement does not allow the anomalous threshold singularity to be present, the space-like form factor does not differ significantly from the unconfined case in the weak confinement limit. Finally (for our purposes) they show that singularities related to this {\em would be} anomalous threshold reappear much farther into the time-like sector of the form factor in the form of poles of oscillating residues.

Although the analysis of Jaffe and Mende is substantially accurate, it introduces several major approximations. For one thing it is inherently nonrelativistic. The lack of relativistic kinematics not only distorts the nature of the time-like singularities but there is also no way to relate their positions to cross channel thresholds. In addition, although their assumed wavefunction is plausible, it is not actually a solution to the Schr\"odinger equation with a known eigenvalue. A precise analysis of singularity positions cannot be made without knowledge of the energy eigenvalue. 

Another area in which clarification is needed is the confinement potential. Jaffe and Mende\cite{jaffe} assumed a potential which flattens to a constant at large separation, reflecting the onset of string breaking. There is on the contrary considerable evidence that the confinement potential continues to rise far beyond the string breaking length. The principal reason is the existence of linear Regge trajectories which most clearly arise from light quarks moving in a linear confinement potential\cite{charlie}.

The present work features a succession of exact analytic solutions of the Klein-Gordon equation. The meson states we consider are thus of the heavy-light type. In each case we use the wavefunction and {\em light} degrees of freedom energy to analytically compute the heavy-light form factor (or Isgur-Wise function) in space and time-like sectors. The singularity structure is easily determined in each case. 

We first briefly review in Section 2 the relativistic form factor expression, the generalized Klein-Gordon equation, and the anomalous threshold singularity. A sequence of solutions are then found and the appropriate form factors are computed. We begin in Section~3 with the spinless relativistic hydrogen atom. The resulting Isgur-Wise form factor for an unconfined meson has a singularity exactly were one would expect the anomalous threshold singularity. In Section~4 we consider a massless quark moving in a scalar confinement potential. The Isgur-Wise (IW) form factor in this case has an essential singularity at the cross channel meson pair threshold. This singularity arises from the confinement of the constituents above and not from any {\em would be} anomalous singularity which is clearly not present in this case. A general solution with linear confinement, massive quark, and Coulomb interaction is established in Section~5. The special case in which a massive, but still light, quark moves in a Coulomb field and pure linear scalar confinement potential is then discussed in Section~6. Here the resulting form factor has only a remnant of a {\em would be} anomalous threshold but still has an essential singularity at the time-like meson threshold. Finally in Section~7 we consider a solution with linear confinement, a Coulomb singularity, and a massive quark. In this heavy quark case the confinement is dynamically unimportant (i.e., weak confinement). The space-like form factor is shown to approach the unconfined anomalous threshold case except now the singularity has been erased. The form factor rises monotonically through the anomalous threshold and has an essential singularity at the meson production threshold. Our summary and conclusions are found in Section~8. 

\section{Wave Equation and Form Factor}

The simplest relativistic wave equation possessing a variety of analytic solutions is the Klein-Gordon equation. For a spinless particle of mass $m$ moving in the field of a Lorentz scalar potential $S(r)$ and a time component vector potential $V(r)$, the generalized Klein-Gordon equation\cite{licht} is
\begin{equation}
{d^2u\over dr^2} = \left[ {\ell(\ell+1)\over r^2} + \left(m^{\vphantom 1} + S(r) \right)^2 - \left(E^{\vphantom 1} - V(r) \right)^2 \right]u \,,
\end{equation}
where $E$ is the light degrees of freedom energy.  
The meson mass $M$ is then found by adding the heavy quark mass $m_Q$,
\begin{equation}
M = m_Q+E \,.  \label{eq:mesonmass}
\end{equation}
In this paper we consider only the ground state (with zero angular momentum) although the generalization to arbitrary orbital and radial states is straightforward. We thus look for solutions $u(r)$ satisfying
\begin{equation}
{d^2u\over dr^2} = \left[  \left(m^{\vphantom 1} + S(r) \right)^2 - \left(E^{\vphantom 1} - V(r) \right)^2 \right]u \label{eq:gen-kg}
\end{equation}
with normalization
\begin{equation}
\int_0^\infty dr | u(r) |^2 = 1 \,.   \label{eq:norm}
\end{equation}

Once a solution $u(r)$ has been found we can compute the Isgur-Wise function $\xi(\omega)$ defined\cite{zalewski} in terms of the overlap between the initial and final states as
\begin{equation}
\xi(\omega) \equiv \sqrt{2\over\omega+1} \left< \Psi(v') \mid \Psi(v) \right>\,,  \label{eq:overlap}
\end{equation}
where $\omega = v_\mu v'^\mu$ is the invariant velocity transfer. The $\sqrt{2\over\omega+1}$ prefactor is required for consistency of the quark model in the heavy quark limit\cite{zalewski,ov-hqet}. This definition is appropriate for a spectator model as shown in Fig.~1a in the heavy quark limit. Lorentz invariance can be exploited to evaluate this overlap integral in the Breit frame, and the result\cite{zalewski,ov-hqet} after completing the angular and time integrations is
\begin{equation}
\xi(\omega) = {2\over\omega+1} \int_0^\infty dr u^2(r) {\sin\alpha r\over \alpha r} \,, \qquad \alpha = 2E\sqrt{\omega-1\over\omega+1} \,.
\label{eq:old4'}
\end{equation}

The velocity transfer $\omega$ is simply related to the invariant momentum transfer $q^2 = (p'_M-p_M)^2$ by
\begin{equation}
q^2 = -2M^2(\omega-1) \,.  \label{eq:q^2}
\end{equation}
The $M\overline M$ threshold at $q^2=4M^2$ corresponds to $\omega=-1$.

Referring to Fig.~1b, it is well known\cite{oehme} that if $M^2>m_Q^2+m^2$ it is possible for $m_Q$ and $m$ to be on mass shell when
\begin{equation}
q_a^2 = {1\over m^2} \left[  m_Q^2 - (M-m)^2 \right] \left[ (M+m)^2 - m_Q^2 \right] \,.
\end{equation}
This defines the anomalous threshold singularity position. In the heavy-light limit we take (\ref{eq:mesonmass}) $M=m_Q+E$, 
with $E \ll m_Q$. If one ignores confinement, the anomalous threshold condition 
  $M^2 - m_Q^2 - m^2 = (m_Q+E)^2 - m_Q^2 - m^2 = 2m_QE+(E^2-m^2)>0$ is satisfied for $E\ll m_Q$. In terms of the light degrees of freedom energy the threshold position is (again in the heavy-light limit)
\begin{equation}
q_a^2 = 4M^2\left(1-{E^2\over m^2}\right)
\end{equation}
or using (\ref{eq:q^2}), in terms of $\omega$,
\begin{equation}
\omega_a = {2E^2\over m^2}-1 \,.  \label{eq:anom-thresh}
\end{equation}

\section{The Unconfined Meson}

We first consider a {\em meson} without confinement. In our case the corresponds in the generalized Klein-Gordon equation (\ref{eq:gen-kg}) to the choice
\begin{equation}
S(r) = 0 \,, \qquad V(r) = -{\kappa\over r} \,,
\end{equation}
which is just the spinless relativistic hydrogen atom satisfying
\begin{equation}
{d^2u\over dr^2} = \left[ m^2 - \left(E+{\kappa\over r}\right)^2\right]u \,.
\label{eq:kg}
\end{equation}
Although the solution for any radial or orbital state can be found by modifying the nonrelativistic hydrogen solution, we will take a simpler approach which works well for the ground state. The solution is assumed to be of the form
\begin{equation}
u(r) = Cr^p e^{-{1\over2}\gamma r} \,.
\end{equation}
By direct substitution into the KG equation (\ref{eq:kg}) this is a solution if we choose
\begin{eqnarray}
\kappa^2 &=& p(1-p) \,,\nonumber\\
\gamma &=& 2m\sqrt{1-p} \,,\label{eq:choose}\\
E &=& m\sqrt{p}  \,. \nonumber
\end{eqnarray}
This reduces to the usual Schr\"odinger result as $p\to 1$ if $\kappa = 
+\sqrt{p\phantom1} \sqrt{1-p}$ (or $2p = 1+\sqrt{1-4\kappa^2}\,$), hence
\begin{equation}
1 \ge p \ge 1/2 \,. \label{eq:p-range}
\end{equation}
For $p<1$ the wavefunction $R=u/r$ will be singular at $r=0$, the well known {\em Coulomb spike}, characteristic of a singular potential with relativistic kinematics. The normalization constant $C$ is determined by
\begin{equation}
1 = C^2 \int_0^\infty dr \, r^{2p} e^{-2m\sqrt{1-p}r} = {C^2\Gamma(2p+1)\over\left(2m\sqrt{1-p}\right)^{2p+1}} \,.
\end{equation}

The IW form factor is then given by (\ref{eq:old4'})
\begin{eqnarray}
& \displaystyle \xi(\omega) = {2C^2\over\omega+1}\; \int_0^\infty dr \, r^{2p-1} e^{-\gamma r} \;{\sin\alpha r\over\alpha} \,, \label{eq:iw} \\
\noalign{\smallskip}
& \displaystyle \alpha = 2m\sqrt p \sqrt{\omega-1\over\omega+1} \,, \qquad
\gamma = 2m\sqrt{1-p} \,.
\end{eqnarray}
At this point it is useful to define a new {\em energy-like} variable $x$ as
\begin{equation}
x \equiv \sqrt{p\vphantom1\over 1-p} \; \sqrt{\omega-1\over\omega+1} \,.
\label{eq:x}
\end{equation}
This new variable maps the $\omega$ upper half plane into the $x$ first quadrant. The $x$ variable will be the natural variable for all of our solutions. In terms of $x$ and a new integration variable,
\begin{equation}
y \equiv \gamma r\,,\qquad \alpha = \gamma x \,,
\end{equation}
the IW function (\ref{eq:iw}) becomes
\begin{eqnarray}
\xi(\omega) &=& {2\over\omega+1}\, {1\over\Gamma(2p+1)} \int_0^\infty dy \, y^{2p-1} e^{-y} \sin xy \nonumber\\
&=& {2\over\omega+1}\, {1\over 4ipx\Gamma(2p)} \int_0^\infty dy \, y^{2p-1} \left(e^{-(y-ix)} - e^{-(y+ix)} \right) \,,
\end{eqnarray}
or\cite{alterform}
\begin{equation}
\xi(\omega) = {2\over\omega+1} {1\over 4ipx} \left[ {1\over(1-ix)^{2p}} - {1\over(1+ix)^{2p}} \right] \,. \label{eq:xi}
\end{equation}

The IW function is singular at only one point along the $\omega$ real axis. For $-1<\omega<1$, $x=\pm i|x|$ and one of the denominators will vanish. In terms of $\omega$ this happens at $\sqrt{p\vphantom1\over 1-p} \sqrt{\omega-1\over\omega+1}=\pm i$ or
\begin{equation}
\omega = 2p-1 \,. \label{eq:singpoint}
\end{equation}
Going back to the location of the anomalous threshold (\ref{eq:anom-thresh}) in our case (\ref{eq:choose}) $E=m\sqrt p$ we have
\begin{equation}
\omega_a = {2E^2\over m^2} - 1 = 2p-1 \,. \label{eq:omega_a}
\end{equation}
Clearly we should identify the IW singularity point (\ref{eq:singpoint}) with an anomalous threshold.

At the strongest allowed Coulomb constant ($p=1/2$) the anomalous singularity is a simple pole at $\omega=0$. For weaker coupling the singularity is a branch point which approaches a double pole for $p=1$ at the physical (space-like) zero recoil point $\omega=1$. Three cases are shown in Fig.~2, with $p=0.5$, 0.7, and 0.9. The physical interpretation is clear. As $p\to1$ the size of the atom, $1/\gamma$, becomes large and the form factor must damp to zero quickly. The space-like form factors for the above three cases are shown in Fig.~3.

The above example, in its nonrelativistic form, was cited by many authors\cite{falk,dominguez} as a counter example to the dispersion result of deRafael and Taron\cite{rafael}, who had claimed an upper limit to the magnitude of the IW slope at the zero recoil point, $\omega=1$. It is clear from Figs.~2 and 3 and from Eq.~(\ref{eq:omega_a}) that as $p$ approaches unity the slope increases without limit. From the expression for the form factor (\ref{eq:xi}) the slope at the zero recoil point is
\begin{equation}
-\xi'(1) = {1\over2} + {p(1+2p)(1+p)\over 6(1-p)} \,. \label{eq:old23'}
\end{equation}
As the Coulomb constant goes to zero ($p\to1$) the slope becomes infinite as previously discussed.

Of course a real meson is confined and since the constituent masses can never be on shell there cannot be an anomalous threshold singularity of this type. As pointed out by several authors\cite{jaffe,falk,dominguez,georgi}, a weakly confined meson should be very similar to an unconfined meson in most respects. In particular, the slope (\ref{eq:old23'}) should be the same in the weakly confined limit and hence would violate the de~Rafael-Taron bound. This {\em would be} or {\em pseudo} anomalous threshold effect will be explicitly demonstrated in Section~7.

\section{Confinement of a Massless Quark: an Essential Singularity}

In this section we consider another instructive and simple example of a Klein-Gordon meson. In this case we ignore the short range interaction entirely, set the quark mass to zero, and assume scalar confinement. The generalized KG Eq.~(\ref{eq:gen-kg}), with $V(r) = m = 0$ and $S(r) = ar$, becomes
\begin{equation}
{d^2u\over dr^2} = \left(a^2 r^2-E^2\right) u\,,
\end{equation}
which is recognizable as the harmonic oscillator equation with ground state solution\cite{charlie}
\begin{eqnarray}
u(r) &=& Cre^{-{1\over2}ar^2} \,, \nonumber\\
E &=& \sqrt{3a} \,,\\
C^2 &=& 4a\sqrt{a\over\pi} \,. \nonumber
\end{eqnarray}
The corresponding IW function (\ref{eq:old4'}) is
\[
\xi(\omega) = {2\over\omega+1} {4a\over2} \sqrt{a\over\pi} \, \sqrt{\omega+1\over 3a(\omega-1)} \int_0^\infty dr\, r e^{-ar^2} \sin\left(2r\sqrt{3a}\,\sqrt{\omega-1\over\omega+1}\;\right)
\]
or
\begin{equation}
\xi(\omega) = {2\over\omega+1} e^{-3\left(\omega-1\over\omega+1\right)} \,.
\label{eq:formfactor}
\end{equation}

One immediately notices that [as in the unconfined Coulomb case (\ref{eq:xi})] the above form factor contains no dimensional parameters. In particular the slope at the zero recoil point ($\omega=1$) is unique and is given by
\begin{equation}
-\xi'(1) = 2 \,.
\end{equation}
This provides a {\em bench mark} for confined mesons. As we will observe later, the addition of an attractive Coulomb interaction will always decrease this value. The form factor (\ref{eq:formfactor}) is depicted in Fig.~4. 
The above IW function (\ref{eq:formfactor}) is remarkably similar to the prediction of the BSW model\cite{neubert} in which the $-3$ in the exponential of (\ref{eq:formfactor}) is replaced by a parameter $\beta=\left<p_T^2\right> / m^2_{\rm spectator}$. 

There is only one singular point of the pure scalar confinement form factor (\ref{eq:formfactor}), and that is an essential singularity at $\omega=-1$. As $\omega$ approaches $-1$ from the positive side the exponent in $\sin\alpha r$ in the IW expression (\ref{eq:old4'}) becomes infinite. Since the meson is confined the large $r$ behavior always is convergent but the form factor necessarily grows exponentially as $\omega\to1(+)$. The other limit $\omega\to1(-)$ involves a rapidly oscillating integrand and there is no singularity. The net result is the observed essential singularity.

Although we have considered a specific model it is clear that the above conclusion is much more general. For a potential which at large $r$ increases as $r^\lambda$; $\lambda\ge0$, the asymptotic KG wavefunction is $u(r) \stackrel{r\to\infty}{\longrightarrow} re^{-(\lambda+1)r}$. For any truly confining potential $\lambda$ is positive definite and the tail of the wavefunction always falls off fast enough so the exponential behavior of the Fourier $\sin\alpha r$ never wins. We then expect an essential singularity to be a universal feature of a meson with confined constituents.

The presence of an essential singularity at $\omega=-1$ was inferred several years ago from perturbative HQET\cite{falk}. In this case the essential singularity is associated with multiple gluon exchanges between comoving heavy quarks. These two singularities are not obviously related since in the latter case it is equivalent to $M\overline M$ interactions, whereas in our case this possibility has not been considered.

The existence of an essential singularity invalidates\cite{falk}  the original dispersion relation proof\cite{rafael} of an upper bound to the magnitude of the IW function slope at $\omega=1$. Later on we will return to the role of confinement in erasing the anomalous threshold singularity. It will turn out that the {\em would be} anomalous threshold and the essential singularity are closely related. 

\section{A More General Solution}

So far we have considered two simple solutions. In Section~3 we examined an unconfined meson bound by Coulomb attraction (i.e., the relativistic hydrogen atom). The corresponding form factor has an anomalous threshold singularity which dictates its behavior in the space-like region. At the other extreme in Section~4 we have constructed a meson consisting of a massless quark confined by a scalar linear potential. The IW form factor (\ref{eq:formfactor}) again has one singularity --- in this case, an essential singularity at the $M\overline M$ threshold ($\omega=-1$). Our aim in this section is to establish a more general solution to the KG equation (\ref{eq:gen-kg}) describing a meson with both a short range Coulomb interaction and linear confinement. The quark should be massive so that one can investigate the limiting cases of weak confinement (large mass) and strong confinement (small mass). We use the wavefunction so obtained to compute (\ref{eq:old4'}) the heavy-light form factor.

Referring back to (\ref{eq:gen-kg}), we will attempt to solve the generalized KG equation with potentials
\begin{equation}
S(r) = far \,, \qquad
V(r) = -{\kappa\over r} + (1-f)ar \,. \label{eq:pots}
\end{equation}
A mixture of scalar and time component vector confinement is assumed with a fraction $f$ scalar confinement. In the nonrelativistic limit the confining potential is $S+V=ar$. Expanding (\ref{eq:gen-kg}) with potentials (\ref{eq:pots}) we have
\begin{equation}
{d^2u\over dr^2} = \left[ -{\kappa^2\over r^2} - {2E\kappa\over r} + \left(m^2-E^2-2\kappa a(f-1)\right) + 2a \left(mf^{\vphantom1} - E(f-1)\right)r + a^2(2f-1)r^2 \right] u \,. \label{eq:expanding}
\end{equation}

As before, we assume the algebraic form for the ground state wavefunction and, by choosing the constants properly,  attempt to satisfy (\ref{eq:expanding}). Our wavefunction choice is motivated by the previous two solutions,
\begin{equation}
u(r) = Cr^pe^{-{1\over2}\gamma r - {1\over2}\beta ar^2} \,. \label{eq:u}
\end{equation}
By differentiation we obtain
\begin{equation}
{d^2u\over dr^2} = \left[ {-p(1-p)\over r^2} - {p\gamma\over r} + \left( {\gamma^2\over 4} - \beta a (2p+1) \right) + \gamma a\beta r + a^2\beta^2 r^2 \right] u(r) \,.  \label{eq:diff}
\end{equation}

For (\ref{eq:diff}) to satisfy (\ref{eq:expanding}) the following relations must be satisfied (assuming $a\neq 0$),
\begin{eqnarray}
\kappa^2 &=& p(1-p) \,, \label{eq:kappa^2}\\
2E\kappa &=& p\gamma \,, \label{eq:2Ekappa}\\
m^2-E^2-2\kappa a(f-1) &=& {\gamma^2\over 4} - \beta a(2p+1) \,,\label{eq:satisfy-c}\\
mf - E(f-1) &=& \gamma \beta/2 \,, \label{eq:mf}\\
2f - 1 &=& \beta^2 \,.
\end{eqnarray}
The first equation (\ref{eq:kappa^2}) is exactly the Coulomb result (\ref{eq:choose}), which defines the short range behavior of the wavefunction. For convenience we parametrize $p$ by an angle $\theta$ defined by the $\theta$ triangle of Fig.~5. Equation (\ref{eq:kappa^2}) can be expressed as
\begin{equation}
\kappa = \sin\theta\cos\theta\,,\qquad p = \cos^2\theta \,. \label{eq:kappa}
\end{equation}
As we noted in Eq.~(\ref{eq:p-range}) of Section~3, $1\ge p \ge 1/2$ (to obtain the Schr\"odinger limit) and hence $0\le\theta\le\pi/4$.
The last relation (\ref{eq:kappa}) can also be parametrized in terms of the angle $\phi$ defined by the $\phi$ triangle of Fig.~5,
\begin{equation}
\beta = \sqrt{2f-1} = f\cos\phi \,. \label{eq:beta-phi}
\end{equation}
With these definitions (\ref{eq:2Ekappa}) becomes
\begin{equation}
{\gamma\over2}\cos\theta = E\sin\theta \label{eq:foo}
\end{equation}
and (\ref{eq:mf}) is 
\begin{equation}
{\gamma\over2} \cos\phi + E\sin\phi = m \,. \label{eq:m}
\end{equation}
From (\ref{eq:beta-phi}) we see $f\ge 1$ and $0\le\phi\le\pi/2$ which in (\ref{eq:m}) implies that $\gamma$ and $E$ are positive. From (\ref{eq:foo}) and (\ref{eq:m}) we find
\begin{eqnarray}
\gamma &=& {2m\sin\theta\over\sin(\theta+\phi)} \,, \label{eq:gamma}\\
E &=& {m\cos\theta\over\sin(\theta+\phi)} \,. \label{eq:E}
\end{eqnarray}
We observe from comparing with (\ref{eq:choose}) that the unconfined (pure Coulomb) limit corresponds to
\begin{equation}
\theta+\phi \stackrel{\rm Coul}{\longrightarrow} \pi/2\,. \label{eq:coul}
\end{equation}

We have not yet considered the solution requirement (\ref{eq:satisfy-c}). From the other requirements we have already defined the observables $E$ and $\gamma$ in terms of the quark mass $m$, Coulomb parameter $p(\theta)$ and the fraction of scalar confinement $f(\phi)$. The existence of (\ref{eq:satisfy-c}) implies that the solution form (\ref{eq:u}) is only valid if the physical parameters are related. Fortunately, the solution still has enough flexibility that we can consider a wide range of mesons and in particular we can interpolate from strong to weak confinement. From (\ref{eq:satisfy-c}), and using the relations (\ref{eq:gamma}) and (\ref{eq:E}) as well as the angles defined in Fig.~5, we obtain
\begin{equation}
{a\over m^2} = {(1-\sin\phi) \cot^2(\theta+\phi)\over 2\cos\phi + \cos(\phi+2\theta)} \,.   \label{eq:a/m^2}
\end{equation}
For a given $a/m^2$ and $\theta$ there is an angle $\phi$ (and hence fraction $f$ of scalar confinement) which must be used for the solution (\ref{eq:u}) to be correct. This fraction lies between $1\le f\le f_{\rm max}$ where from the zero of the denominator of (\ref{eq:a/m^2}) we have
\begin{equation}
f_{\rm max} = {1+8p\over 4p(1-p)}\,. \label{eq:old44'}
\end{equation}
Two types of mesons will be of particular interest: a)~strong confinement ($m^2\alt a$) with pure scalar confinement, and b)~weak confinement when $m^2\gg a$. These limiting situations will be more fully explored in the next two sections. 

Once the wavefunction and light degrees of freedom energy are known, the IW form factor can be constructed using (\ref{eq:old4'}). Our solution will involve the {\em Weber function}\cite{gradstein} (often known as the {\em parabolic cylinder function}) defined by
\begin{equation}
D_{-q}(z) \equiv {e^{-z^2\over4}\over\Gamma(q)} \int_0^\infty dy \, y^{q-1} e^{-yz-{1\over2}y^2} \,. \label{eq:weberfunct}
\end{equation}
For $q>0$ it is evident that $D_{-q}(z)$ is an analytic function for all finite $z$.

For our wavefunction (\ref{eq:u}) the normalization is found by
\begin{equation}
1 = C^2 \int_0^\infty dr \, r^{2p} e^{-\gamma r-\beta ar^2} = C^2 {\Gamma(2p+1)\over (2\beta a)^{p+{1\over2}}} e^{\lambda^2\over4} D_{-(2p+1)}(\lambda) \,,
\end{equation}
where we have defined
\begin{equation}
\lambda\equiv \gamma/\sqrt{2\beta a} \,.  \label{eq:lambda}
\end{equation}
The IW function is determined by (\ref{eq:old4'}) as
\begin{equation}
\xi(\omega) = {2\over\omega+1} C^2 \int_0^\infty dr \, r^{2p-1} e^{-\gamma r - \beta a r^2} \;{\sin\alpha r\over \alpha} \,.
\end{equation}
The Fourier variable $\alpha$ can be written as
\begin{equation}
\alpha = 2E\sqrt{\omega-1\over\omega+1} = \left( 2E\sqrt{1-p\over p}\right) \left(\sqrt{p\vphantom1\over 1-p}\; \sqrt{\omega-1\over\omega+1}\right) \,.
\end{equation}
As before, (\ref{eq:x}), we define 
\begin{equation}
x \equiv \sqrt{p\vphantom1\over 1-p} \; \sqrt{\omega-1\over\omega+1}
\label{eq:x-again}
\end{equation}
and using (\ref{eq:foo}) we see that
\begin{equation}
\alpha = \gamma x
\end{equation}
and
\begin{equation}
{\alpha\over\sqrt{2\beta a}} = \lambda x \,.
\end{equation}
in terms of Weber functions the IW function is
\begin{equation}
\xi(\omega) = {2\over\omega+1}\; {e^{-\lambda^2x^2/4}\over 4ip\lambda x D_{(-2p+1)}(\lambda)} \left[ e^{-ix\lambda^2/2} D_{-2p}\left(\lambda^{\vphantom1}(1-ix)\right) - e^{ix\lambda^2/2} D_{-2p} \left(\lambda^{\vphantom1}(1+ix)\right) \right]\,,  \label{eq:iw-weber}
\end{equation}
where $\lambda$ and $x$ were defined in (\ref{eq:lambda}) and (\ref{eq:x-again}) respectively. This result will be used extensively in the subsequent two sections.

\section{A Strongly Confined Meson}

The general result of the preceding section is first applied to a meson with relatively small mass moving in a pure scalar confinement and Coulomb attraction. Setting $\phi=0$ (for pure scalar confinement) in the general solution, (\ref{eq:gamma}), (\ref{eq:E}), and (\ref{eq:a/m^2}) then yield
\begin{eqnarray}
\gamma &=& 2m \,,\nonumber\\
E &=& m\sqrt{p\over 1-p} \,,\label{eq:yield}\\
{m^2\over a} &=& (1+2p)(1-p)/p \,.\nonumber
\end{eqnarray}
For $1\ge p\ge 1/2$ we note that the allowed quark mass range is
\begin{equation}
0\le m\le \sqrt{2a} \,.
\end{equation}
From (\ref{eq:yield}) we also observe that
\begin{eqnarray}
\lambda^2 &=& {\gamma^2\over2a} = 2(1+2p) (1-p) / p \,,\\
E^2 &=& a(1+2p) \,.
\end{eqnarray}
When $p\to 1$, then $E\to\sqrt{3a}$,  $m=0$ and we recover the result discussed in Section~4. One can verify from (\ref{eq:iw-weber}) that (\ref{eq:formfactor}) results if one takes the correct limits $\lambda^2 x^2 \stackrel{p\to1}{\longrightarrow} 6\left(\omega-1\over\omega+1\right)$ and $\lambda^2 x\stackrel{p\to1}{\longrightarrow}0$.

In Fig.~6 we plot the slope at the zero recoil point as a function of the Coulomb parameter $p$. We see that the slope decreases in magnitude as $p$ decreases (i.e., the attraction increases). One should keep in mind that as $p$ changes the meson mass changes as required by (\ref{eq:yield}). For the strongest allowed Coulomb constant the slope has decreased monotonically to $-\xi'(1)\simeq 0.71$. At the smaller values of $p$ (corresponding to the largest Coulomb constant) we are starting to see the remnant of a {\em would be} anomalous singularity. As $p$ increases toward unity the unconfined meson would expand without limit. With confinement the size of the meson is roughly $1/\sqrt a$ and decreasing the Coulomb constant does not increase the meson size.

\section{The Weakly Confined Meson}

The limit $m\gg\sqrt a$, known as weak confinement, is particularly interesting. This limit of QCD has been discussed in general\cite{georgi} and was the primary focus of the analysis of Jaffe and Mende\cite{jaffe}. In this limit the quark moves nonrelativistically  and the interaction can be described perturbatively. The meson dynamics are thus nearly independent of the existence of large distance confinement. Nevertheless, the weak confinement limit is not well defined\cite{durand} in the sense that it is not analytic in the string tension $a$. For a negative string tension, no matter how small, the wavefunction is not normalizable.

To look at our solution in the $m^2\gg a$ limit we note from (\ref{eq:a/m^2}) that
\begin{equation}
\theta+\phi \stackrel{m^2\gg a}{\longrightarrow} \pi/2 \,,
\end{equation}
which we had already noted in (\ref{eq:coul}) was the pure Coulomb limit. If we take $\phi={\pi\over2}-\theta$ we see that this is equivalent to
\begin{equation}
f = {1\over1-\sqrt p} \,, \qquad \beta = \sqrt{2f-1} = {\cos\phi\over 1-\sin\phi} \,.
\end{equation}
The IW function (\ref{eq:iw-weber}) depends on the parameter $\lambda$,
\begin{equation}
\lambda = {\gamma\over\sqrt{2\beta a}} = {\sqrt{2\,}m\over\sqrt a} \, {\sin\theta\over\sin(\theta+\phi)} \, \sqrt{1-\sin\phi\over\cos\phi} \,.
\label{eq:lambdagain}
\end{equation}
In the limit of interest in this section, $m\gg \sqrt a$ and $\phi={\pi\over2}-\theta$, (\ref{eq:lambdagain}) becomes
\begin{equation}
\lambda^2 \stackrel{m^2\gg a}{\longrightarrow} {2m^2\over a} \sqrt{1-p} \left(1-\sqrt p\right) \,.  \label{eq:becomes}
\end{equation}
When $p\neq 1$, the weak confinement limit is given by $\lambda\gg1$.

The weak confinement limit is now easy to analyze. In the definition of the Weber function (\ref{eq:weberfunct}) if Re\,$z\gg1$ then the linear exponential causes the integral to converge before the quadratic term has had appreciable effect. We may then approximate,
\begin{equation}
D_{-q}(z) \stackrel{{\rm Re}\,z\gg 1}{\longrightarrow} {e^{-z^2/4}\over z^q} \,.  \label{eq:approx}
\end{equation}

\noindent\underline{\bf Space-like region}

For $\omega>1$, $x$ is real and the approximation (\ref{eq:approx}) is valid for all terms of the general IW expression (\ref{eq:iw-weber}). A brief calculation shows that the unconfined Coulomb form factor (\ref{eq:xi}) is recovered,
\begin{equation}
\xi(\omega) \stackrel{m^2\gg a}{\longrightarrow} {2\over\omega+1} \, {1\over4ipx} \left[ (1-ix)^{-2p} - (1+ix)^{-2p} \right] \,.
\end{equation}
The accuracy becomes greater the heavier the quark. This is the {\em would be} anomalous threshold effect.

\noindent\underline{\bf Anomalous threshold region}

As $\omega$ moves into the time-like region ($\omega<1$), the defined (\ref{eq:x-again}) quantity $x$ becomes pure imaginary. To be definite we choose the $x\to +i|x|$ branch but this is arbitrary and it would not change the result if we had taken the other sign.
When $x=i$ the approximation (\ref{eq:approx}) cannot be made for the second term $D_{-2p}(\lambda(1+ix))$ where the argument vanishes. The point $x=i$ was the anomalous threshold position in the unconfined case (\ref{eq:xi}). Now however the Weber function still converges due to the quadratic term and
\begin{equation}
D_{-q}(0) = 2^{{q\over2}-1} \Gamma(q/2) / \Gamma(q) \,. \label{eq:result}
\end{equation}
Using the still valid approximations (\ref{eq:approx}) for $D_{-(2p+1)}(\lambda)$ and $D_{-2p}(2\lambda)$ and the result (\ref{eq:result}) in the general IW expression (\ref{eq:iw-weber}) we find for $m\gg\sqrt a$ that at the anomalous singularity position
\begin{equation}
\xi(\omega_a=2p-1) \simeq {2^{2p-3} \Gamma(p)(1-p)^{p/2} \left(1-\sqrt p\right)^p\over p^2 \Gamma(2p)} \left(m^2\over a\right)^p \,.
\end{equation}
In Fig.~7 the weakly confined and unconfined form factors are shown near the anomalous threshold singularity ($\omega=0.4$ for $p=0.7$). The unconfined meson of course becomes infinite but the confined form factor remains finite. We show the cases $m/\sqrt a= 5$, 10, and 20. The space-like region is shown in Fig.~8 for the same choices of $m/\sqrt a$. The approach to the unconfined case is evident.

\noindent\underline{\bf The essential singularity}

As $\omega$ decreases from $\omega_a=2p-1$ (or $|x|>1$) the linear exponential in $D_{-2p}(\lambda(1-|x|))$ has a positive slope. The quadratic term of (\ref{eq:weberfunct}) is then critical for convergence. The IW function rises exponentially as $\omega$ approaches $-1$ because of the large contribution of the linear term. At $\omega=-1$ the variable $x$ jumps (on our branch) from $+i\infty$ to $+\infty$ and the IW function has a zero for $\omega\to 1(-)$. This is an essential singularity at $\omega=-1$. As we discussed earlier in Section~4, this is a general conclusion within our basic assumptions. Any confining potential will give a wavefunction whose tail falls off faster than a linear exponential. Both the Coulombic wavefunction and the Fourier transform generate linear exponentials and thus any confining potential will both erase the anomalous threshold singularity and give rise to an essential singularity at $\omega=-1$. Jaffe and Mende\cite{jaffe} assumed that confinement gave a linear exponential tail with a larger constant then the Coulomb part. The result in this case was to generate singularities shifted to a sequence of more time-like points.

Finally, it should be emphasized that the elimination of the anomalous threshold singularity is not achieved by pushing it off the physical sheet. This would leave a peak of finite height near $\omega_a$, whereas the IW function is actually monotonic. The singularity vanishes in a non-analytic manner, leaving no obvious trace at $\omega_a$.

\section{Summary and Conclusions}

In this paper we have sought to investigate more precisely the relation of the confinement of constituents on the analytic nature of the heavy-light form factor. To this end we have used the information contained in the Fourier transform of the meson wavefunction. Our results are basically a relativistic version of the work of Jaffe and Mende\cite{jaffe} with closer attention given to the physical nature of the wavefunction tail  implied by the confining potential. Our main result is that for a confining (i.e., rising) potential the large distance wavefunction tail erases any anomalous threshold singularity and simultaneously induces an essential singularity at the cross channel {\em normal} threshold.

We examine the analytic structure by use of a relativistic form factor definition dictated by heavy quark symmetry. The mesons we consider are also in the heavy-light limit and are solutions to the generalized Klein-Gordon equation (\ref{eq:gen-kg}).

First we have considered an unconfined meson bound by Coulomb attraction. The Isgur-Wise form factor is controlled by an anomalous threshold singularity. We next examine the solution for a massless quark confined by a linear scalar potential but with no short range interaction. The associated Isgur-Wise form factor now has an essential singularity at $\omega=-1$ (which is the time-like normal threshold). It was observed that this singularity follows generally for any confining potential.

Our central argument uses a new solution to the Klein-Gordon equation in which a quark of mass $m$ moves in a central Coulomb field and a superposition of scalar and time component vector linear confining potentials. Our analytic solution interpolates between the two previous cases, allowing form factors corresponding to strong and weak confinement to be investigated for their analytic properties. In the weak confinement limit we show that the space-like form factor sector is identical to that produced from the unconfined anomalous threshold singularity except the singularity has been erased. This is the {\em would be} anomalous threshold of Jaffe and Mende\cite{jaffe}. Even in this weakly confined case there will always be an essential singularity at $\omega=-1$. The same confinement mechanism which erases the anomalous singularity generates the essential singularity.

One way to summarize some of our results is a direct calculation of the meson size. The proper relativistic measure of the size is the slope of the IW function at the zero recoil point. From (\ref{eq:old4'}) one finds by direct differentiation\cite{zalewski,ov-hqet}
\begin{equation}
-\xi'(1) = {1\over2} + {1\over3} E^2 \left< r^2 \right> \,.
\end{equation}
From our general solution of Section~5 we find that
\begin{equation}
-\xi'(1) = {1\over2} + {p(p+1) (2p+1)\over 6(1-p)}\lambda^2 {D_{-(2p+3)}(\lambda)\over D_{-(2p+1)}(\lambda)} \,,  \label{eq:xi'}
\end{equation}
where $\lambda=\gamma/\sqrt{2\beta a}$. There are two interesting limits where the slope can be analytically evaluated.

\begin{itemize}

\item[a)] Weak confinement ($m^2\gg a$ or $\lambda\gg1$).\\
In this limit we use (\ref{eq:approx}) to show that ${D_{-(2p+3)}(\lambda)/ D_{-(2p+1)}(\lambda)}\simeq 1/\lambda^2$ and hence (\ref{eq:xi'}) becomes
\begin{equation}
-\xi'\simeq{1\over2} + {p(p+1)(2p+1)\over6(1-p)} \,,
\end{equation}
which is exactly the unconfined Coulomb limit (\ref{eq:old23'}). This is an example of the {\em would be} anomalous threshold effect. In the weak confinement limit the meson size is the same as the unconfined Coulomb meson. The latter is entirely due to an anomalous threshold singularity below the $M\overline M$ threshold. The weakly confined meson has no singularity (in our model) below the $M\overline M$ threshold.

\item[b)] Strong confinement ($m^2\alt a$).\\
For pure scalar confinement our solution has $m^2\le 2a$ and $\lambda^2=2(2p+1)(1-p)/p$ and (\ref{eq:xi'}) becomes
\begin{equation}
-\xi'(1) = {1\over2} + {(p+1)(2p+1)^2\over 3} \; {D_{-(2p+3)}(\lambda)\over D_{-(2p+1)}(\lambda)} \,. \label{eq:xi'-strong}
\end{equation}
The size of the strongly confined meson reaches a maximum (see Fig.~6) at $p=1$ which corresponds to a massless quark. In this case $\lambda=0$ and by (\ref{eq:result}) we have $D_{-(2p+3)}(0) / D_{-(2p+1)}(0)=1/4$ and (\ref{eq:xi'-strong}) becomes
\begin{equation}
-\xi'(1) = 2 \,.
\end{equation}
In this strongly confined case the meson size does not increase without limit as $p\to1$ and the {\em would be} effect loses its significance.

\end{itemize}

Of course there are additional singularities of the form factor associated with interaction between the heavy quark pair. Although we have not considered them here, they can also generate an essential singularity at $\omega=-1$\cite{falk}. It is useful to point out that the $m_Q\overline m_Q$ bound or resonant states play an intrinsically different role than the heavy-light interaction in terms of their effect on the Isgur-Wise form factor.  The two-particle Klein-Gordon equation can be solved by the same technique as the one-particle equation and one can see that for a Coulomb interaction the singularity closest to the space-like regime is at $\omega=1-2p_2$ where here $\kappa^2=p_2(1-p_2)$. For weak attraction $p_2\to1$ and the singularity retreats to the normal threshold $\omega=-1$ (in the limit $m_Q\gg m$). This should be contrasted to the heavy-light case discussed in Section~3, where for weak attraction the anomalous threshold approaches $\omega=1$. Even for the strongest possible $m_Q\overline m_Q$ interaction the singularity is always located at $\omega\leq 0$. 

\section*{Acknowledgments}

I would like to thank Sini\v sa Veseli, Charles Goebel, and Gustavo Burdman for helpful conversations. 
This research was supported in part by the U.S.~Department of Energy under Grant No.~DE-FG02-95ER40896 and in part by the University of Wisconsin Research Committee with funds granted by the Wisconsin Alumni Research Foundation.


\newpage
\section*{Figures}

\begin{itemize}

\item[Fig.~1:] (a) Spectator model for heavy-light meson transition from velocity $v$ to $v'$ through the interaction of a current and the heavy quark.\\
(b) Same model as in (a) but drawn in the time-like current region to show the possibility of an anomalous threshold.

\item[Fig.~2:] Unconfined meson (pure Coulomb) Isgur-Wise function (\ref{eq:xi}) for various Coulomb constants $\kappa = \sqrt{p(1-p)}$. The anomalous threshold singularity occurs at $\omega_a=2p-1$. For weak attraction ($p\to1$) the meson becomes large and the singularity approaches the zero-recoil point $\omega=1$. 

\item[Fig.~3:] Unconfined (Coulombic) Isgur-Wise form factor in the space-like region for Coulombic attraction varying from strongest ($p=0.5$) to weak ($p=0.9$). The large slope at $\omega=1$ in the $p=0.9$ case reflects the nearby singularity as shown in Fig.~2.

\item[Fig.~4:] Massless light quark confined by a linear scalar potential. The Isgur-Wise function (\ref{eq:formfactor}) has a slope $-\xi'(1)=2$ at the zero recoil point and rises to become infinite at $\omega=-1$. The function is very small for $\omega<-1$.

\item[Fig.~5:] Angles parametrizing the Coulomb interaction ($\theta$) confinement ($\phi$).

\item[Fig.~6:] Strongly confined meson Isgur-Wise slope at zero recoil. At the strongest Coulomb interaction ($p=0.5$) the slope approaches the unconfined (pure Coulomb) slope.

\item[Fig.~7:] Weak confinement limit as $m/\sqrt a$ becomes large. The Isgur-Wise form factor is shown for $0.4<\omega<1$. An unconfined meson has an anomalous threshold singularity at $\omega=0.4$ in the case of $p=0.7$ (see Fig.~2). A variety of quark masses are shown, none of which are singular at $\omega=0.4$. These form factors approximate the unconfined case with increasing accuracy as $m/\sqrt a$ increases in the region $\omega>0.4$.

\item[Fig.~8:] Weak confinement approach to the unconfined space-like form factor as $m/\sqrt a$ increases. In this limit, $m/\sqrt a\gg 1$, a confined meson and an unconfined meson having the same Coulomb constant become equal. The case shown is for $p=0.7$. Referring to Fig.~7 the anomalous threshold singularity is not present for the confined meson. This is the {\em would be} anomalous effect.

\end{itemize}

\newpage
\pagestyle{empty}

\begin{center}

\hspace{0in}\epsfysize=7in\epsffile{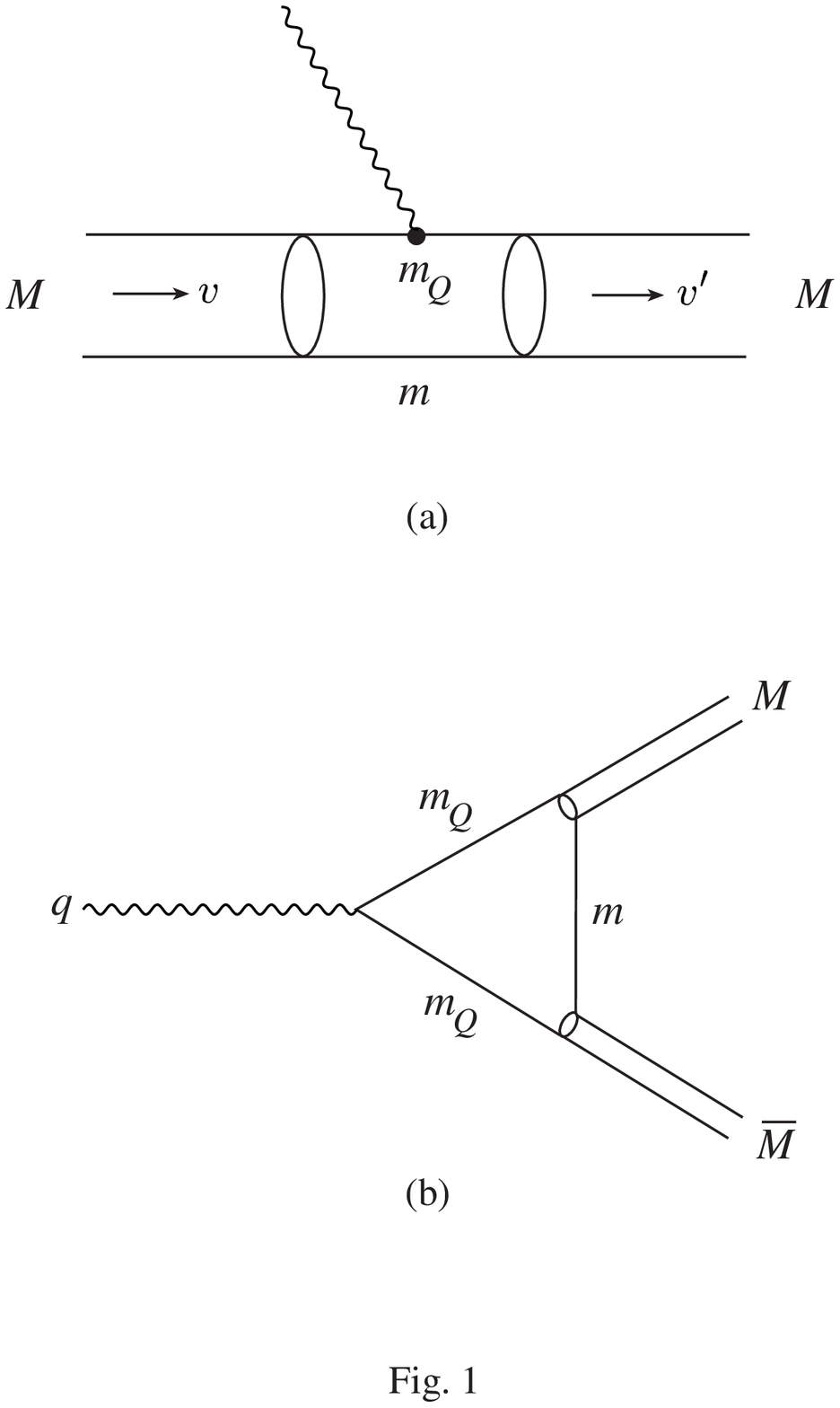}

\newpage
\hspace{0in}\epsfysize=8in\epsffile{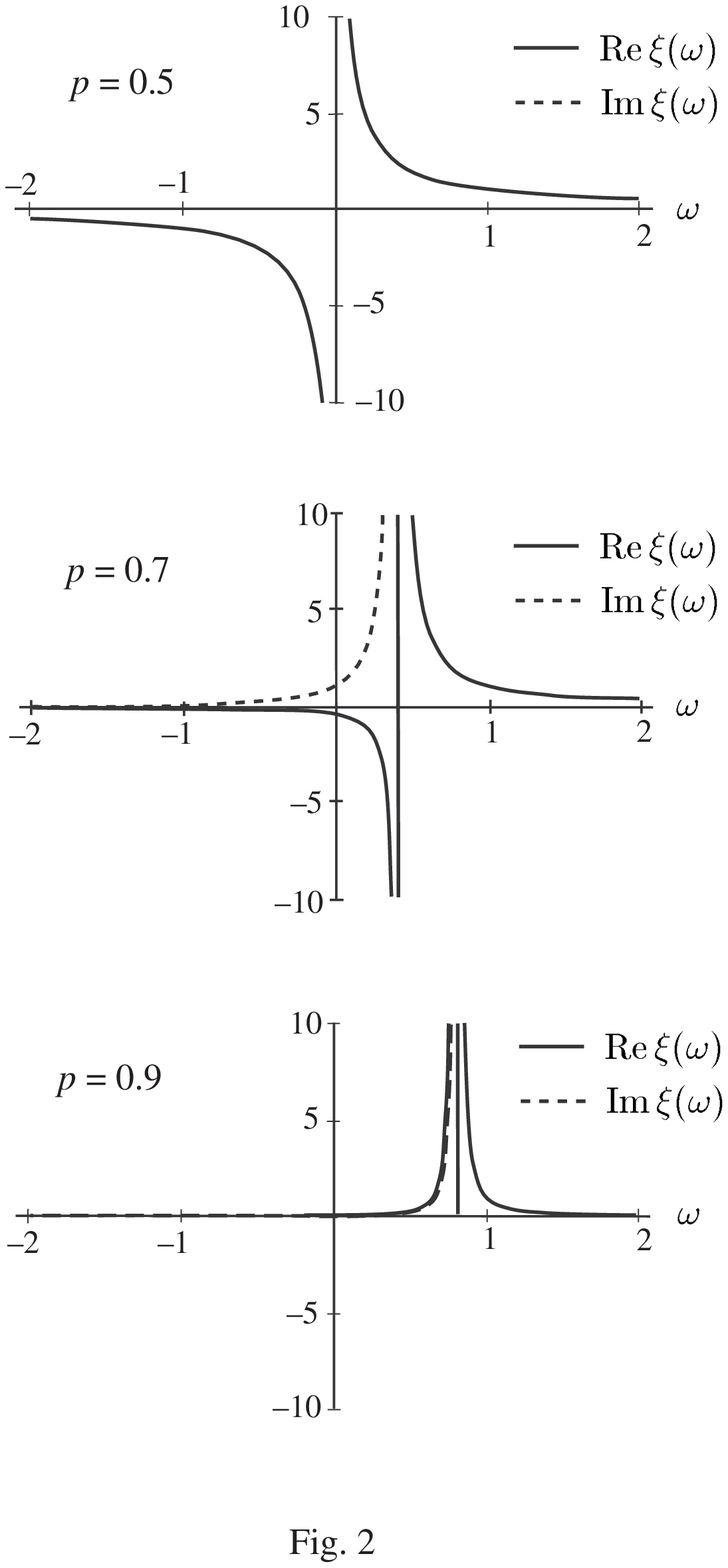}

\newpage
\vspace*{2in}

\hspace{0in}\epsffile{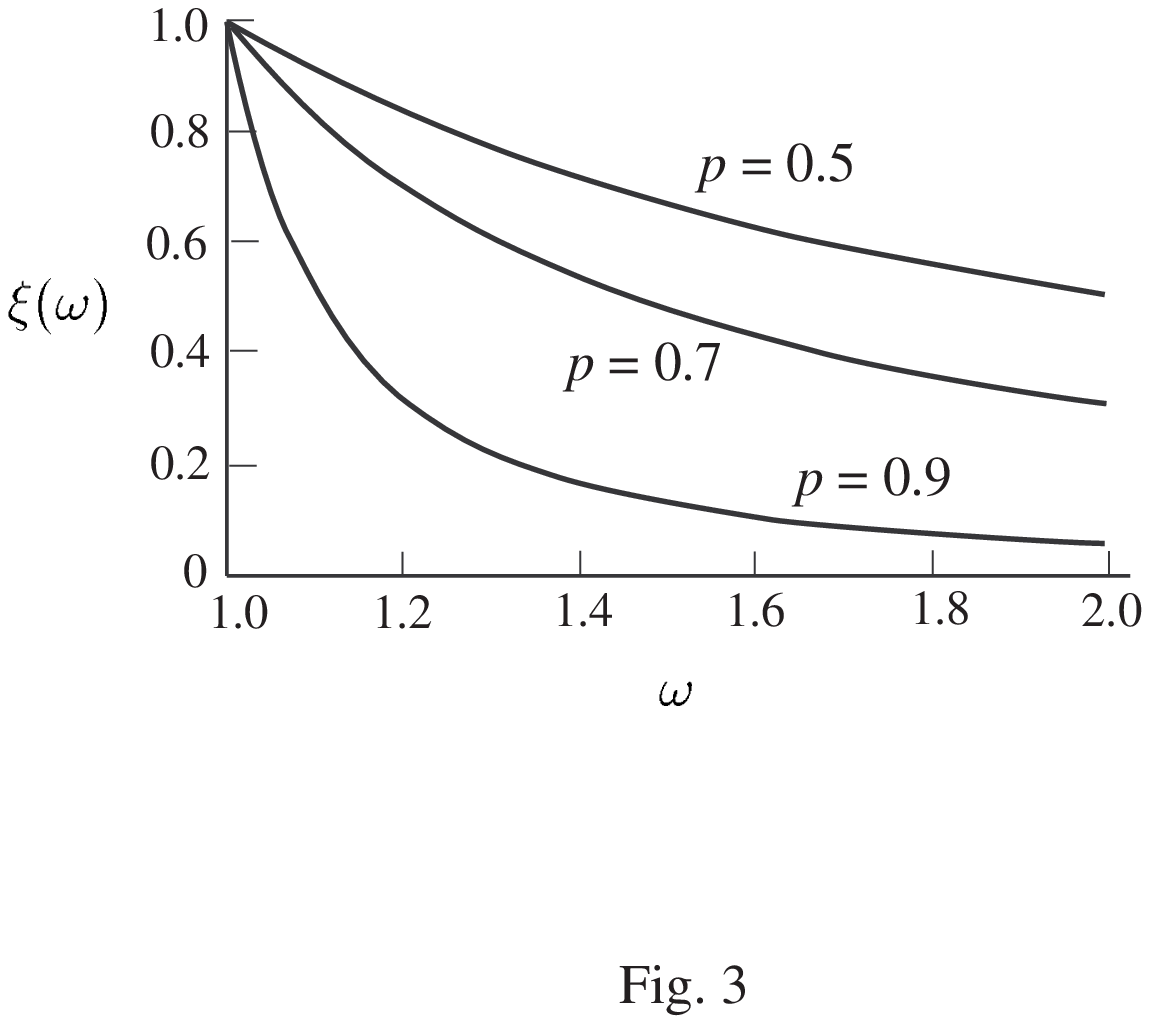}

\newpage
\vspace*{2in}

\hspace{0in}\epsffile{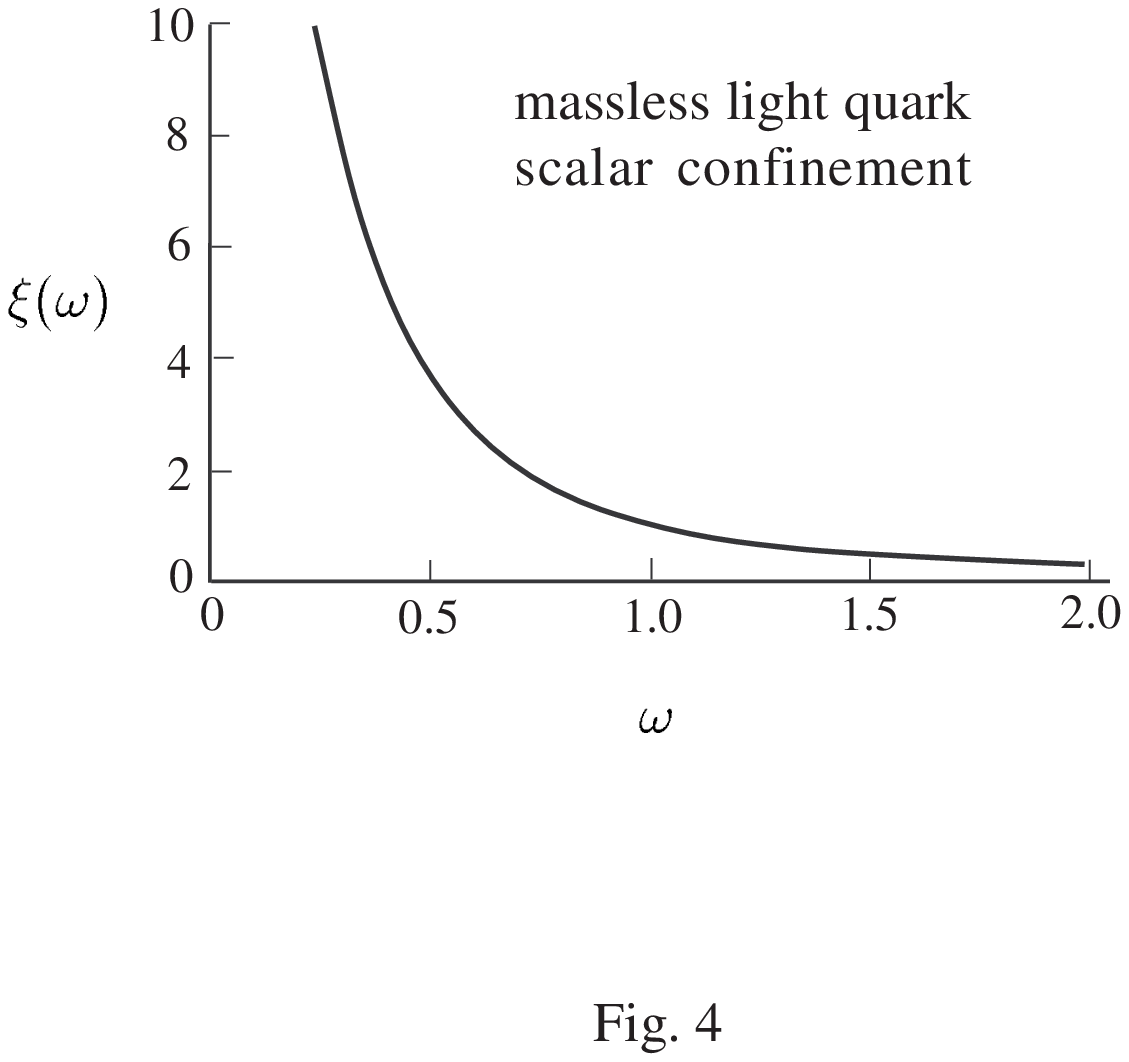}

\newpage
\vspace*{1in}

\hspace{0in}\epsffile{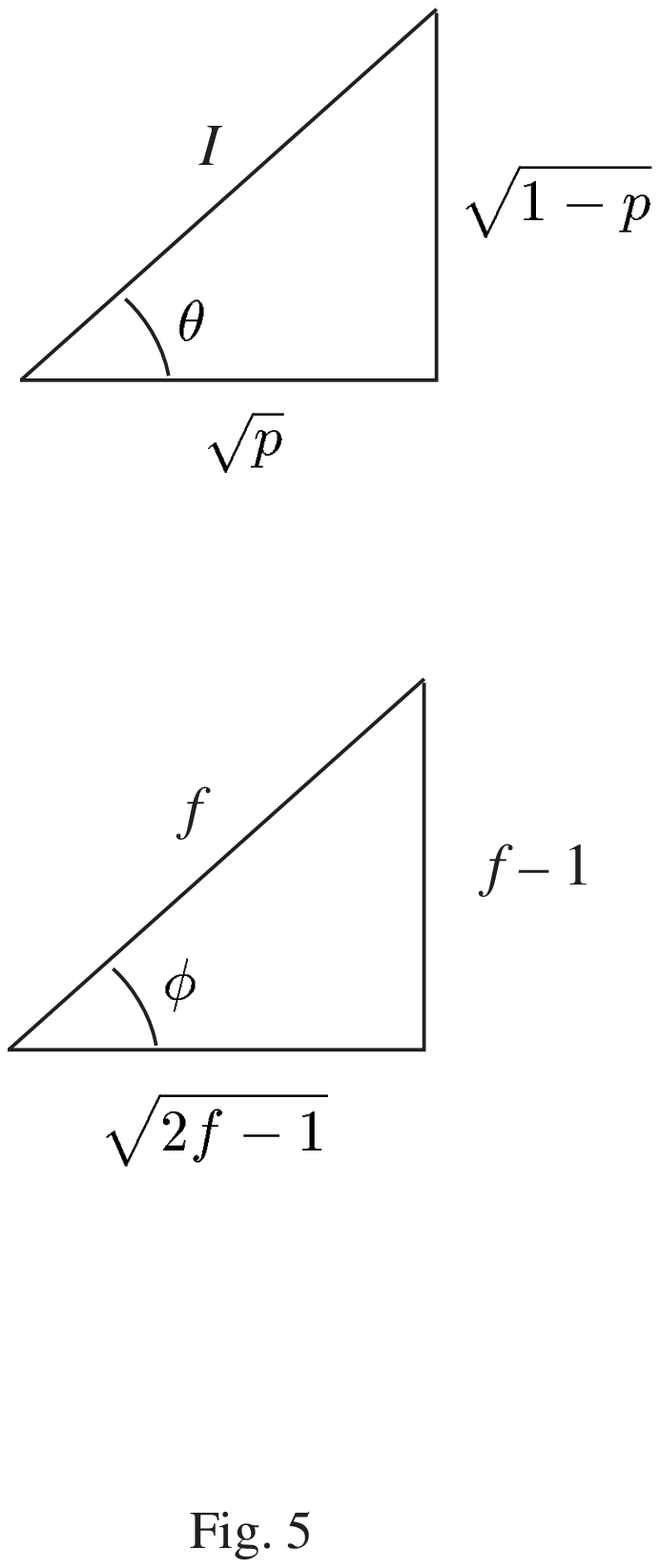}

\newpage
\vspace*{2in}

\hspace{0in}\epsffile{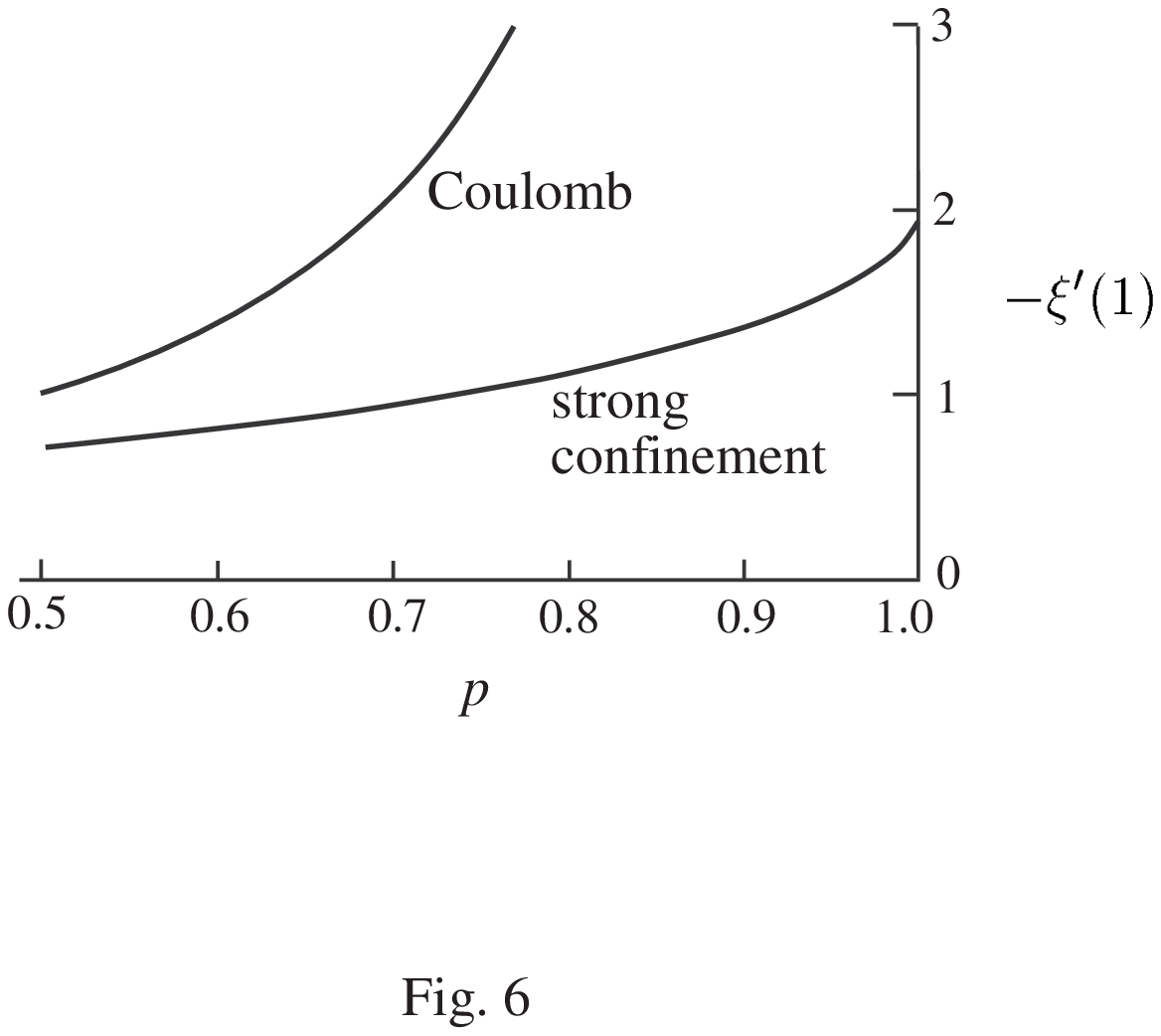}

\newpage
\vspace*{2in}

\hspace{0in}\epsffile{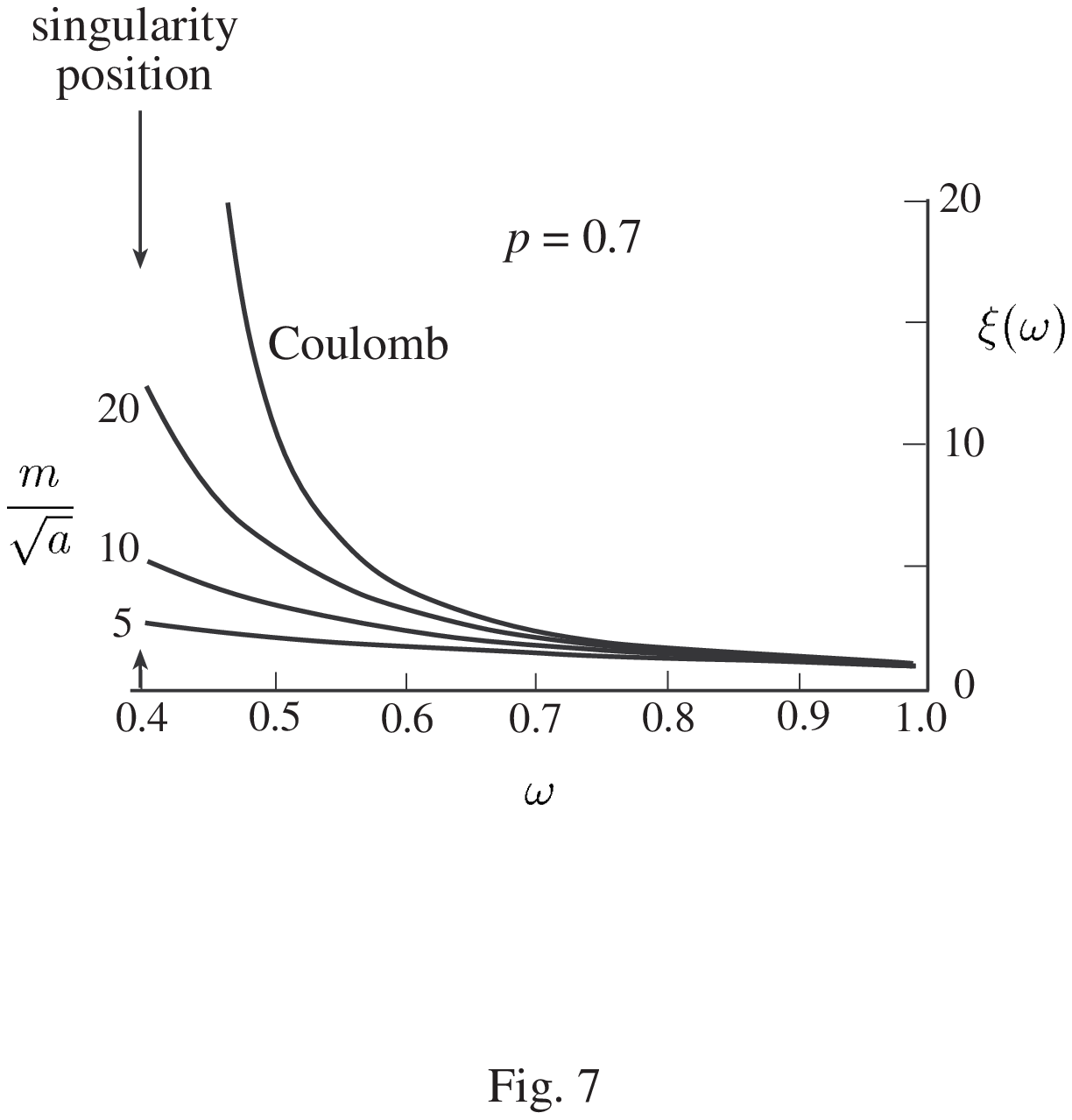}

\newpage
\vspace*{2in}

\hspace{0in}\epsffile{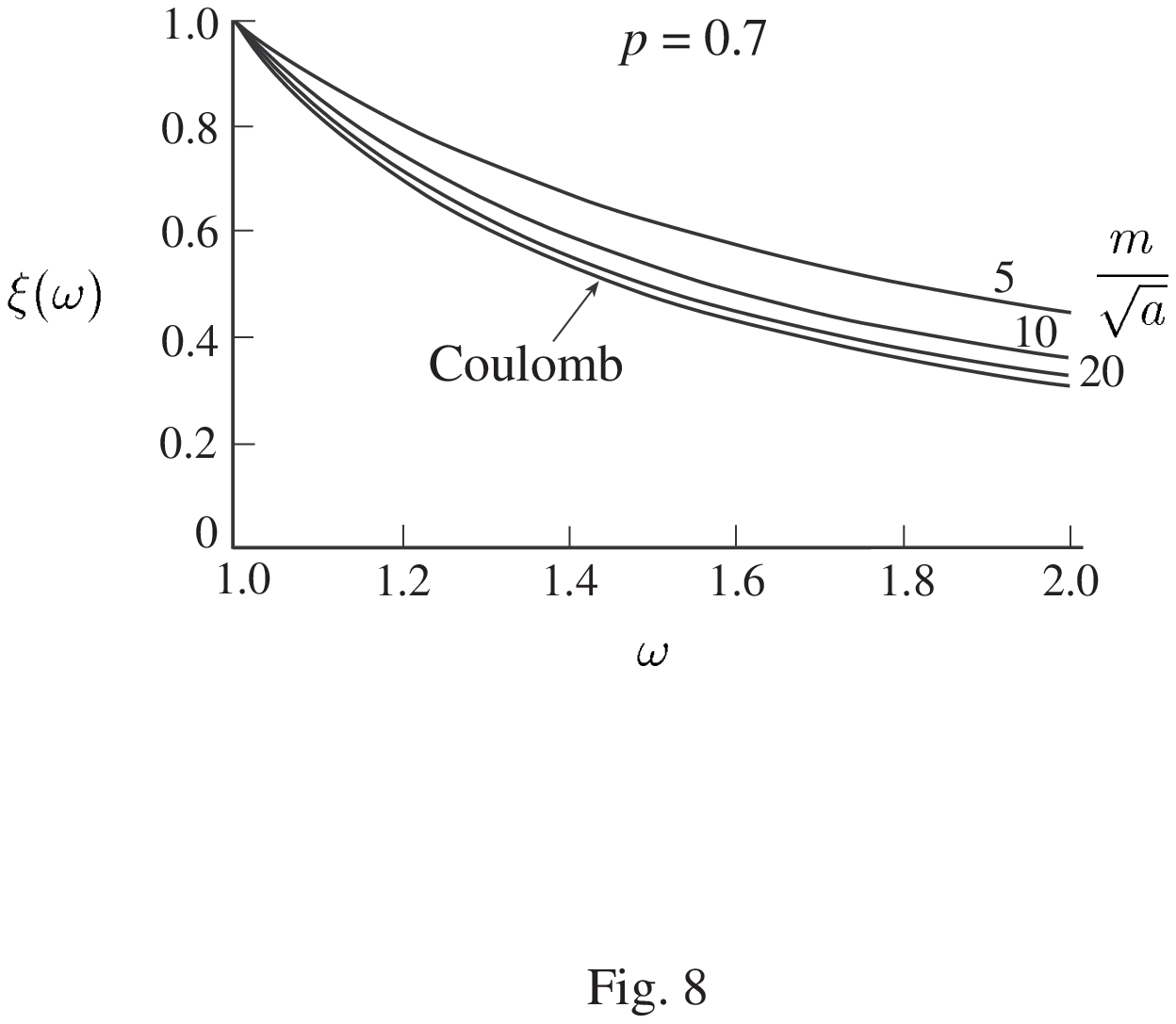}

\end{center}

\begin{thebibliography}{99}
\bibitem{alterform}
A useful alternative form, especially for $\omega>1$, is obtained using the polar form $1+ix=re^{i\theta}$, which yields $\xi(\omega) = {2\over\omega+1} \, {1\over 2px} \, {1\over\left(1+x^2\right)^p} \sin\left[2p\tan^{-1}x\right]$.
\end{thebibliography}
\end{document}